\documentclass{article}
\usepackage{spconf}
\usepackage{cite}
\usepackage{url}
\usepackage{amsmath,amssymb}
\usepackage{booktabs}
\usepackage{array}
\usepackage{graphicx}
\usepackage{float}
\floatstyle{ruled}
\newfloat{algorithm}{tbp}{loa}
\floatname{algorithm}{Algorithm}


\title{DISTILL LOCALLY, SCHEDULE GLOBALLY:\\
FLOW MAPS FOR FEW-STEP TEXT-TO-SPEECH}

\name{\shortstack{Yentl Collin 
\qquad Evan Dufraisse \\
Amr Mohamed
\qquad Amine Khelif Khelif 
\qquad Dani Bouch
\qquad Guokan Shang}}

\address{Institute of Foundation Models (IFM), MBZUAI}

\begin{document}
\ninept
\maketitle
\raggedbottom

\vspace{-7pt}

\begin{abstract}
Flow-matching text-to-speech (TTS) models achieve high synthesis quality but require many neural function evaluations (NFEs) to integrate their generative trajectories. Recent few-step flow-map distillation approaches for TTS construct targets from numerically integrated teacher trajectories, creating a trade-off between target accuracy and training cost. We propose Local Flow-Map Distillation (LFMD), which adapts Eulerian Map Distillation to conditional TTS and avoids teacher trajectory integration during target construction. For inference, we derive a sampling schedule (TD-DP) from teacher dynamics and consistency of the learned maps, with a single cost graph supporting multiple NFE budgets without external audio-metric evaluation. Because scheduling offers no flexibility at one NFE, we refine this regime with alignment-aware temporal self-distillation using soft-DTW. Across Seed-TTS and LibriSpeech-PC, LFMD improves low-NFE synthesis over a matched integral-distillation baseline. On Seed-TTS, the refined student reaches 1.80\% WER with 1-NFE, compared with 1.76\% for its 32-NFE teacher.

\end{abstract}


\vspace{-6pt}

\section{Introduction}
\label{sec:intro}

Non-autoregressive diffusion and flow-matching models have enabled high-quality zero-shot text-to-speech (TTS) synthesis. Models such as E2 TTS~\cite{eskimez2024e2tts} and F5-TTS~\cite{chen2025f5tts} generate acoustic representations by transporting noise through an ordinary differential equation (ODE). Their inference cost, however, scales with the number of NFEs required by numerical integration.

Few-step generation can reduce this cost through improved sampling schedules~\cite{zheng2025fastf5tts} or model distillation. Flow-map methods~\cite{boffi2024flowmapmatching,geng2025meanflow,sabour2025alignyourflow,boffi2025flowmaps} are particularly attractive because they learn direct transport between arbitrary departure and arrival times, enabling flexible inference budgets. Most have been developed primarily for image generation. Recent TTS flow-map methods include SplitMeanFlow~\cite{guo2025splitmeanflow}, IntMeanFlow~\cite{wang2025intmeanflow}, and DSFlow~\cite{lin2026dsflow}. IntMeanFlow and DSFlow's mean-velocity branch construct average-velocity targets from numerically integrated teacher displacements, avoiding the Jacobian--vector products (JVPs) required by differential objectives. IntMeanFlow highlights the memory and operator-compatibility challenges of such JVPs for TTS. These trajectory-integrated targets, however, require many teacher evaluations and depend on trajectory discretization, creating a trade-off between target accuracy and training cost.

We propose Local Flow-Map Distillation (LFMD), adapting Eulerian flow-map distillation~\cite{sabour2025alignyourflow} to conditional speech generation. Rather than constructing targets from integrated teacher trajectories, LFMD uses a local differential identity requiring only the teacher velocity at the departure state. We reduce the memory overhead of JVP-based objectives through detached forward-mode target construction, without retaining an unnecessary reverse-mode graph. A progressive interval curriculum and short-interval teacher-velocity anchor further stabilize the learning of long transports. LFMD therefore avoids numerical teacher integration over the target interval. In our experiments, it improves low-NFE synthesis over our matched integral-distillation baseline, particularly at one and two NFEs, while reducing sensitivity to inference scheduling.

Despite this reduced sensitivity, inference scheduling remains important at low NFE. Existing schedules range from empirical pruning~\cite{zheng2025fastf5tts} to metric-driven search such as O3S~\cite{wang2025intmeanflow}, which requires repeated synthesis and a separate search for each NFE budget. We introduce teacher-derived dynamic programming (TD-DP), which uses local flow-map consistency and map composition to construct a single interval-cost graph reusable across inference budgets without an external audio metric.

Scheduling improves multi-step generation, but at one NFE there is no intermediate step to optimize. We therefore refine the lowest-NFE regime through temporal self-distillation from a stronger frozen four-step LFMD model. Soft-DTW~\cite{cuturi2017softdtw,blondel2021dtw} accommodates local timing differences between multistep targets and lower-step predictions and outperforms frame-wise self-distillation in our experiments. The refined 1-NFE student reaches 1.80\% WER on Seed-TTS, close to the 32-NFE teacher's 1.76\%, while remaining below the teacher in SIM-o (.607 vs.\ .667) and UTMOS (3.41 vs.\ 3.76).

Our contributions are:
\begin{itemize}
\setlength{\itemsep}{0pt}
\setlength{\parsep}{0pt}
\setlength{\topsep}{2pt}
\item A memory-efficient and stabilized adaptation of Eulerian flow-map distillation to TTS that avoids teacher trajectory integration.
\item A Teacher-Derived inference scheduling (TD-DP), which reuses one flow-map consistency graph to select schedules across NFE budgets without external audio metrics.
\item Alignment-aware temporal self-distillation that further improves one-step synthesis without changing the inference-time architecture.
\end{itemize}

\vspace{-10pt}

\section{Related Work}
\label{sec:related_work}

\subsection{Flow Maps and Integral Velocity Distillation}

Few-step generation has been explored through consistency models~\cite{song2023consistency}, shortcut models~\cite{frans2024shortcut}, Flow Map Matching~\cite{boffi2024flowmapmatching}, MeanFlow~\cite{geng2025meanflow}, Align Your Flow~\cite{sabour2025alignyourflow}, and flow-map self-distillation~\cite{boffi2025flowmaps}, mostly for image generation. These approaches learn finite-time transport, either toward a common end point or between specified departure and arrival times. Recent TTS flow-map distillation methods include SplitMeanFlow~\cite{guo2025splitmeanflow}, IntMeanFlow~\cite{wang2025intmeanflow}, and DSFlow~\cite{lin2026dsflow}. IntMeanFlow and DSFlow's mean-velocity branch supervise average velocity using teacher displacements divided by interval duration; DSFlow additionally matches teacher and student rollout end points. Integral targets depend on the teacher ODE solver, sampling grid, and NFE budget. More teacher steps increase training cost, while coarse integration can reduce target accuracy over long intervals. Our IntMeanFlow (IMF) reproduction yields comparable results to the original paper~\cite{wang2025intmeanflow} and exhibits the same cost-quality trade-off. Target discretization and scheduling can particularly affect low-NFE synthesis, which requires long transports. LFMD instead uses local teacher velocity, avoiding trajectory integration during target construction.

\subsection{Inference-Time Step Selection}

Step selection is widely studied for accelerating flow-based generation.
EPSS~\cite{zheng2025fastf5tts} derives a training-free schedule from F5-TTS
trajectories: it retains closely spaced steps in the initially curved,
noisy portion and prunes steps from the straighter later portion. It can
also be combined with Sway sampling~\cite{chen2025f5tts}. IntMeanFlow's
O3S~\cite{wang2025intmeanflow} instead adjusts step positions through
repeated synthesis and scoring with an auxiliary speech model. Search is
repeated for each NFE budget and depends on the metric, evaluator, and
development set. In image generation, GITS~\cite{chen2024gits} and Optimal
Stepsize Distillation~\cite{pei2025optimalsteps} use trajectory information
and dynamic programming. With a teacher already available from distillation,
we introduce TD-DP to use its local dynamics and student map composition
as interval costs. A single graph can support a schedule search across NFE budgets
without the need for external speech evaluator.

\subsection{Metric Optimization and Adversarial Fine-Tuning}

DMOSpeech~\cite{li2025dmospeech} and DMOSpeech~2~\cite{li2026dmospeech2} optimize perceptual metrics using auxiliary speech models, while adversarial approaches such as Align Your Flow~\cite{sabour2025alignyourflow} and APTTS~\cite{yoon2025aptts} rely on discriminators. We instead refine low-NFE synthesis through self-distillation within our flow-map model, using stronger multistep predictions from a frozen LFMD model as targets for lower-step generation. Motivated by prior use of soft-DTW in TTS to handle temporal mismatch~\cite{elias2021paralleltacotron2,yi2022softspeech}, we use an alignment-aware self-distillation loss to accommodate local timing differences between multistep targets and lower-step predictions. In our experiments, this formulation substantially outperforms frame-wise self-distillation at one NFE, while requiring neither an external evaluator nor a discriminator.

\vspace{-8pt}

\section{Methodology}
\label{sec:methodology}

\subsection{Conditional Flow Maps}
\label{sec:conditional_maps}

We first train an F5-TTS model from scratch, then freeze and use it as teacher
for the distillation. Conditioned on text and an audio prompt $c$, it transports
Gaussian noise to a mel-spectrogram according to
\begin{equation}
\frac{d x_\tau}{d\tau}=v_T(x_\tau,\tau,c),\qquad \tau\in[0,1].
\label{eq:teacher_ode}
\end{equation}
Here, $x_\tau$ is the intermediate acoustic state and $v_T$ is the teacher's
learned flow-matching velocity, including classifier-free guidance (CFG)~\cite{ho2022cfg}.
In our formalism, $t=0$ corresponds to noise and $t=1$ to data. The teacher flow map is defined as:
\begin{equation}
\Phi_T^{t\to s}(x_t,c)
=x_t+\int_t^s v_T(x_\tau,\tau,c)\,d\tau,\qquad t\leq s.
\label{eq:teacher_map}
\end{equation}
We approximate it with an Euler-form neural map:
\begin{equation}
f_\theta(x_t,t,s,c)=x_t+(s-t)u_\theta(x_t,t,s,c),
\label{eq:euler_map}
\end{equation}
where $u_\theta$, the student, is trained to predict the teacher's interval-averaged
velocity. This parameterization enforces the
identity $f_\theta(x_t,t,t,c)=x_t$; consistency with the teacher additionally
requires $u_\theta(x_t,t,s,c)\to v_T(x_t,t,c)$ as $s\to t$. The student
absorbs the CFG during distillation and uses one conditional evaluation per
interval.

\subsection{Local Flow-Map Distillation}
\label{sec:local_distillation}

For $t<s$, let $\bar u_T$ denote the integral in
Eq.~\eqref{eq:teacher_map} divided by $s-t$. Differentiating along a teacher
trajectory, with arrival time $s$ fixed, gives
$\frac{d}{dt}[(s-t)\bar u_T]=-v_T$, hence
\begin{equation}
\bar u_T=v_T+(s-t)\frac{d\bar u_T}{dt}.
\label{eq:eulerian_identity}
\end{equation}
This is the departure-time flow-map
identity~\cite{geng2025meanflow,sabour2025alignyourflow,boffi2025flowmaps}.
Throughout, $d/dt$ follows the teacher field $dx_t/dt=v_T$, with $s$ and
$c$ fixed. Thus $\frac{d}{dt}\Phi_T^{t\to s}(x_t,c)=0$: moving the departure
state along its teacher trajectory preserves the end point. LFMD enforces
this local constraint to learn the flow map without a teacher rollout over $[t,s]$.

\textbf{Training states and curriculum.}
We initialize the student from the teacher and sample interpolated states
$x_t=(1-t)\epsilon+t x_1$, where $x_1$ is a training mel-spectrogram and
$\epsilon\sim\mathcal N(0,I)$. At each iteration $k$, we bound by $h_k$ the length of the sampled interval 
$\Delta=s-t$, which increases linearly to the full
horizon. Another coefficient $\beta_k$, the derivative correction, simultaneously ramps from zero to $0.95$. Together, these curricula allow
the teacher-initialized model to learn short transports before long jumps. The stabilization components in LFMD were selected through targeted early-training sweeps over the interval curriculum, derivative-correction schedule, and short-interval anchor. We retained the reported configuration because it consistently improved optimization stability, particularly when training included long transport intervals.

\textbf{Detached local target.}
A forward-mode JVP along $(v_T,1)$ computes the total derivative
$\dot u=du_\theta/dt=\partial_tu_\theta+(\nabla_xu_\theta)v_T$.
This direction uses the teacher velocity at the sampled state. We construct
a detached target $y$ and regress $u_\theta$ toward it:
\begin{gather}
y=\operatorname{sg}\!\left[u_\theta-
\mathcal N(u_\theta-v_T-\beta_k\Delta\dot u)\right],\nonumber\\
\mathcal L_{\mathrm{LFMD}}=\mathbb E\!\left[\|u_\theta-y\|_M^2\right].
\label{eq:local_loss}
\end{gather}
Here, $\operatorname{sg}$ stops gradients and
$\mathcal N(r)=r/(\operatorname{RMS}_M(r)+0.1)$ normalizes the residual.
The masked norm averages over generated frames and mel channels.
A short-interval anchor $\lambda_a\alpha(\Delta)\|u_\theta-v_T\|_M^2$
uses a weight $\alpha$ that decays to zero while interval length increases.

Algorithm~\ref{alg:lfmd} separates the trainable prediction from its auxiliary target. Reverse-mode recording is disabled for the teacher and JVP branch, while forward-mode differentiation remains active. Backpropagation therefore traverses only the prediction branch. In our PyTorch implementation, this reduces peak training memory by approximately a factor of six compared with an implementation that constructs and retains a reverse-mode graph for the JVP branch until the target is detached.

\begin{algorithm}[t]
\caption{One LFMD training update}
\label{alg:lfmd}
\fontsize{9}{9}\selectfont
\textbf{Inputs:} frozen teacher $v_T$, student $u_\theta$, interval bound
$h_k$, tangent weight $\beta_k$, anchor weight $\lambda_a$.\par\smallskip
\begin{tabular}{@{}r@{\hspace{0.4em}}p{0.93\columnwidth}@{}}
1 & Sample data $(x_1,c)$, noise $\epsilon$, and $0<\Delta\leq h_k$.\\
2 & Sample $t\in[0,1-\Delta]$; set $s=t+\Delta$ and
    $x_t=(1-t)\epsilon+t x_1$.\\[2pt]
3 & \textbf{Prediction (gradients enabled):}\\
4 & \quad $\widehat u\leftarrow u_\theta(x_t,t,s,c)$.\\[2pt]
5 & \textbf{Target (reverse-mode off; forward-mode on):}\\
6 & \quad $v\leftarrow v_T(x_t,t,c)$.\\
7 & \quad Fix $g(x,\tau)=u_\theta(x,\tau,s,c)$.\\
8 & \quad $(\widetilde u,\dot u)\leftarrow
    \operatorname{JVP}(g;(x_t,t);(v,1))$.\\
9 & \quad $r\leftarrow\widetilde u-v-\beta_k\Delta\dot u$.\\
10 & \quad $y\leftarrow\operatorname{sg}[\widetilde u-\mathcal N(r)]$.\\[2pt]
11 & \textbf{Loss and update:}\\
12 & $\ell\leftarrow\|\widehat u-y\|_M^2+
    \lambda_a\alpha(\Delta)\|\widehat u-v\|_M^2$.\\
13 & Backpropagate through $\widehat u$ and update $\theta$.
\end{tabular}
\par\smallskip
The JVP returns the primal $\widetilde u=g(x_t,t)$ and tangent
$\dot u=\frac{d}{dt}u_\theta(x_t,t,s,c)$ along the teacher field, with $s,c$ fixed.
\end{algorithm}

\subsection{Teacher-Derived Step Scheduling}
\label{sec:td_dp_method}

EPSS empirically prunes a grid, while O3S optimizes it through repeated
audio-metric evaluation. Teacher-derived dynamic programming (TD-DP)
instead uses the teacher and student to
construct a directed interval-cost graph on $0=t_0<\cdots<t_J=1$.
Calibration states use the same flow-matching interpolation as training.

\textbf{Local teacher consistency.}
The LFMD identity also defines an interval cost. For an edge $(i,j)$,
we hold arrival time $t_j$ fixed and measure the map's total derivative
along the teacher field:
\begin{equation}
C_T(i,j)=\mathbb E\!\left[
\left\|\left.\frac{d}{dt}f_\theta(x_t,t,t_j,c)\right|_{t=t_i}\right\|_M^2
\right].
\label{eq:teacher_cost}
\end{equation}
By Eq.~\eqref{eq:euler_map}, this derivative equals
$v_T-u_\theta+(t_j-t)\,du_\theta/dt$. Its negative is the undamped LFMD
residual. The cost therefore reuses the same JVP, without tangent damping
or residual normalization, to measure local end point consistency.

\textbf{Composition consistency.}
Exact flow maps satisfy the semigroup property: direct transport equals
two successive transports through any intermediate time $t_i<m<t_j$.
We measure this discrepancy by composing the student map explicitly:
\begin{equation}
C_C(i,j)=\mathbb E\!\big[\|f_\theta(x_{t_i},t_i,t_j,c)
\!-\!f_\theta(f_\theta(x_{t_i},t_i,m,c),m,t_j,c)\|_M^2\big].
\label{eq:composition_cost}
\end{equation}
For each edge, $m$ is chosen deterministically as the midpoint between
$t_i$ and $t_j$ in logit time, with end point clipping for numerical
stability. Costs are averaged over calibration utterances and noise draws
and normalized by their positive medians. As $C_T$ and $C_C$ capture complementary local and compositional consistency, respectively, we use
$C=\widehat C_T+\widehat C_C$, which performed better in validation than using either cost alone.

\textbf{Dynamic programming.}
For $K$ evaluations, the additive schedule solves
\begin{equation}
\min_{0=i_0<\cdots<i_K=J}\;
\sum_{k=0}^{K-1}C(i_k,i_{k+1}).
\label{eq:dp_schedule}
\end{equation}
The same graph
supports a separate path for each NFE budget, without recomputing edge
costs or using an external speech evaluator.
\subsection{Temporal Self-Distillation for Low-NFE Synthesis}
\label{sec:dtw_method}

With TD-DP, four-step LFMD outperforms its 32-step teacher in WER
and approaches it in speaker similarity and predicted quality.
One-step synthesis still leaves a gap. Self-distillation trains lower-NFE
predictions toward the stronger four-step model's output distribution,
using four target evaluations instead of a 32-step teacher rollout. Let
$G_\theta^{(K)}$ compose $K$ maps from time zero to one, and freeze a copy
$\theta_0$ of the distilled model. For $K\in\{1,2,3\}$, shared noise $x_0$,
and conditioning $c$, the student output and target are
\begin{equation}
S=G_\theta^{(K)}(x_0,c),\qquad
Y=\operatorname{sg}[G_{\theta_0}^{(4)}(x_0,c)].
\label{eq:temporal_target}
\end{equation}

Framewise regression penalizes small timing shifts between otherwise
similar acoustic sequences. We instead use band-constrained
soft-DTW~\cite{cuturi2017softdtw}. After cropping the generated span and
pooling in time, let $X=P(S)$ and $Z=P(Y)$ have length $N$. We denote by
$s_{\gamma,b}(X,Z)$ the soft-DTW cost computed from mean squared distances
between mel frames over monotone alignment paths constrained by
$|i-j|\leq b$, with $\gamma>0$ controlling the soft minimum over paths.
With the shorthand $s_{XZ}=s_{\gamma,b}(X,Z)$, the self-comparison
correction~\cite{blondel2021dtw} is
\begin{equation}
\mathcal D(X,Z)=\big(s_{XZ}-\tfrac12s_{XX}-\tfrac12s_{ZZ}\big)/N.
\label{eq:dtw_correction}
\end{equation}

We pool every four frames, set $b=4$ and $\gamma=0.1$, and
combine temporal alignment with full-resolution and source regularization:
\begingroup
\thinmuskip=0.5mu
\medmuskip=0.5mu
\thickmuskip=1mu
\begin{equation}
\mathcal L_{\mathrm{FT}}
=
\mathbb E\big[
\mathcal D(P(S),P(Y))
+0.1\left\|S-Y\right\|_M^2
+0.1\left\|S-S_0\right\|_M^2
\big],
\label{eq:finetuning_loss}
\end{equation}
\endgroup
where $S_0=\operatorname{sg}[G_{\theta_0}^{(K)}(x_0,c)]$ is the frozen
model's output at the student's budget. The two regularizers preserve
acoustic detail and limit drift, respectively. Frozen models are needed
only during training; inference keeps the same architecture and NFE budget.

\vspace{-5pt}

\section{Experiments}
\label{sec:experiments}

\begin{table*}[t]
\centering
\caption{Teacher and few-step student results. WER is in \%, SIM denotes SIM-o, and bold marks the best student score for each dataset among models matched in training data, size, and NFE (ties included). DTW denotes temporal self-distillation. 
}
\label{tab:progression}
\fontsize{9}{9.9}\selectfont
\setlength{\tabcolsep}{0.7pt}
\renewcommand{\arraystretch}{0.89}
\newcommand{\metrichead}[2]{{\fontsize{9}{9}\selectfont #1$#2$}}

\begin{tabular*}{\textwidth}{@{\extracolsep{\fill}}l@{\hspace{4pt}}l@{\hspace{4pt}}l@{\hspace{5pt}}
*{3}{w{c}{25.5pt}}@{\hspace{5pt}}
*{3}{w{c}{25.5pt}}@{\hspace{5pt}}
*{3}{w{c}{25.5pt}}@{\hspace{5pt}}
*{3}{w{c}{25.5pt}}@{}}

\toprule

\multicolumn{3}{@{}l}{\textbf{F5-TTS teachers}} &
\multicolumn{12}{c}{32 NFEs} \\
\cmidrule(lr){4-15}

Training & Size & Evaluation &
\multicolumn{4}{c}{\metrichead{WER}{\downarrow}} &
\multicolumn{4}{c}{\metrichead{SIM}{\uparrow}} &
\multicolumn{4}{c}{\metrichead{UTMOS}{\uparrow}} \\
\cmidrule(lr){4-7}
\cmidrule(lr){8-11}
\cmidrule(lr){12-15}

Emilia & Small  & Seed-TTS      & \multicolumn{4}{c}{1.88} & \multicolumn{4}{c}{.625} & \multicolumn{4}{c}{3.82} \\
Emilia & Base   & Seed-TTS      & \multicolumn{4}{c}{1.76} & \multicolumn{4}{c}{.667} & \multicolumn{4}{c}{3.76} \\
Emilia & Medium & Seed-TTS      & \multicolumn{4}{c}{1.92} & \multicolumn{4}{c}{.682} & \multicolumn{4}{c}{3.72} \\
LibriTTS & Small & LibriSpeech-PC & \multicolumn{4}{c}{2.23} & \multicolumn{4}{c}{.582} & \multicolumn{4}{c}{4.07} \\

\midrule

& & &
\multicolumn{3}{c}{1 NFE} &
\multicolumn{3}{c}{2 NFEs} &
\multicolumn{3}{c}{3 NFEs} &
\multicolumn{3}{c}{4 NFEs} \\
\cmidrule(lr){4-6}
\cmidrule(lr){7-9}
\cmidrule(lr){10-12}
\cmidrule(lr){13-15}

Training & Size & Variant &
\metrichead{WER}{\downarrow} & \metrichead{SIM}{\uparrow} & \metrichead{UTMOS}{\uparrow} &
\metrichead{WER}{\downarrow} & \metrichead{SIM}{\uparrow} & \metrichead{UTMOS}{\uparrow} &
\metrichead{WER}{\downarrow} & \metrichead{SIM}{\uparrow} & \metrichead{UTMOS}{\uparrow} &
\metrichead{WER}{\downarrow} & \metrichead{SIM}{\uparrow} & \metrichead{UTMOS}{\uparrow} \\

\midrule

\multicolumn{15}{@{}l}{\textbf{LFMD (ours)}} \\
\multicolumn{15}{@{}l}{\emph{Seed-TTS test-en}} \\

Emilia & Small & EPSS
& 7.00 & .522 & 2.18
& 5.05 & .557 & 3.00
& 2.59 & .587 & 3.50
& 1.96 & .588 & 3.57 \\

Emilia & Base & EPSS
& 5.49 & .558 & 2.82
& 3.98 & .597 & 3.27
& 2.22 & .619 & 3.46
& 1.83 & .633 & 3.55 \\

Emilia & Base & O3S
& 5.49 & .558 & 2.82
& 1.75 & .616 & 3.46
& 1.75 & .624 & 3.56
& 1.74 & .634 & 3.57 \\

Emilia & Base & TD-DP
& 5.49 & .558 & 2.82
& 1.76 & .618 & 3.46
& 1.72 & .625 & 3.58
& 1.59 & .637 & 3.70 \\

Emilia & Base & TD-DP + DTW
& \textbf{1.80} & \textbf{.607} & \textbf{3.41}
& \textbf{1.68} & \textbf{.626} & \textbf{3.58}
& \textbf{1.61} & .633 & \textbf{3.59}
& \textbf{1.59} & .637 & 3.70 \\

\addlinespace[1pt]
\multicolumn{15}{@{}l}{\emph{LibriSpeech-PC test-clean}} \\

LibriTTS & Small & EPSS
& \textbf{6.85} & \textbf{.510} & \textbf{2.28}
& 4.65 & .529 & 3.11
& 2.81 & .568 & 3.79
& 2.33 & .576 & 3.93 \\

LibriTTS & Small & O3S
& \textbf{6.85} & \textbf{.510} & \textbf{2.28}
& 2.52 & \textbf{.573} & 3.75
& 2.33 & \textbf{.583} & 3.75
& 2.08 & .580 & 3.94 \\

LibriTTS & Small & TD-DP
& \textbf{6.85} & \textbf{.510} & \textbf{2.28}
& \textbf{2.40} & .569 & \textbf{3.79}
& \textbf{2.10} & .579 & \textbf{3.94}
& \textbf{2.01} & .581 & 4.03 \\

\midrule

\multicolumn{15}{@{}l}{\textbf{IMF (Reproduced; 16-step Sway-sampled teacher integration)}} \\
\multicolumn{15}{@{}l}{\emph{Seed-TTS test-en}} \\

Emilia & Base & EPSS
& 7.77 & .515 & 1.98
& 4.84 & .608 & 3.01
& 2.38 & .638 & 3.51
& 1.78 & .651 & \textbf{3.73} \\

Emilia & Base & O3S
& 7.77 & .515 & 1.98
& 2.14 & .618 & 3.34
& 1.98 & \textbf{.648} & 3.51
& 1.68 & \textbf{.655} & \textbf{3.73} \\

\addlinespace[1pt]
\multicolumn{15}{@{}l}{\emph{LibriSpeech-PC test-clean}} \\

LibriTTS & Small & EPSS
& 8.08 & .508 & 2.18
& 4.93 & .557 & 3.39
& 4.15 & .564 & 3.65
& 2.40 & .580 & 4.01 \\

LibriTTS & Small & O3S
& 8.08 & .508 & 2.18
& 2.82 & .570 & 3.72
& 2.33 & .574 & 3.70
& 2.23 & \textbf{.584} & \textbf{4.05} \\
\midrule
\multicolumn{15}{@{}l}{
\textbf{Model-sampling RTF ($\times 10^{-2}$)}
\hspace{0.35em}\emph{Emilia, AMD MI210; LFMD and IMF have matched inference cost.}
} \\[2pt]

& \multicolumn{12}{c}{Student} &
\multicolumn{2}{c}{Teacher} \\
\cmidrule(lr){2-13}
\cmidrule(l){14-15}

Size &
\multicolumn{3}{c}{1 NFE} &
\multicolumn{3}{c}{2 NFEs} &
\multicolumn{3}{c}{3 NFEs} &
\multicolumn{3}{c}{4 NFEs} &
\multicolumn{2}{c}{32 NFEs} \\
\cmidrule(lr){2-4}
\cmidrule(lr){5-7}
\cmidrule(lr){8-10}
\cmidrule(lr){11-13}
\cmidrule(l){14-15}

Small &
\multicolumn{3}{c}{1.32} &
\multicolumn{3}{c}{1.67} &
\multicolumn{3}{c}{2.04} &
\multicolumn{3}{c}{2.34} &
\multicolumn{2}{c}{12.69} \\

Base &
\multicolumn{3}{c}{1.63} &
\multicolumn{3}{c}{2.08} &
\multicolumn{3}{c}{2.58} &
\multicolumn{3}{c}{3.06} &
\multicolumn{2}{c}{20.33} \\

\bottomrule
\end{tabular*}
\end{table*}

\subsection{Experimental Setup}
\label{sec:setup}

\textbf{Models and data.}
We train F5-TTS~\cite{chen2025f5tts} Small (160M), Base (350M), and Medium (600M) teachers from
scratch, on 95k hours of
English and Chinese Emilia speech~\cite{he2024emilia}. We also train a Small
teacher on the 585-hour LibriTTS corpus~\cite{zen2019libritts}. Distillation
uses Small and Base models; Medium is an additional teacher reference.
Vocos~\cite{siuzdak2023vocos} converts mel-spectrograms to 24-kHz waveforms.

Because no public IntMeanFlow implementation is available, we reproduce it under settings matched to LFMD and obtain results consistent with the original paper. For the main comparison, IMF training targets are constructed using 16-step Sway-sampled teacher integration. Following IntMeanFlow, we use identity/zero initialization for the departure/arrival-time projection and matched conditioning scales. Both students absorb teacher CFG~\cite{ho2022cfg} and require one conditional evaluation per inference interval.

\textbf{Evaluation and sampling.}
Emilia-trained models are evaluated on Seed-TTS test-en~\cite{anastassiou2024seedtts};
LibriTTS-trained models use LibriSpeech-PC test-clean under F5-TTS's
cross-sentence protocol. We report Whisper large-v3 WER~\cite{radford2022whisper},
WavLM-based SIM-o~\cite{chen2022wavlm}, and UTMOS~\cite{saeki2022utmos},
averaged over three generation seeds. Students use one to four NFEs with
EPSS~\cite{zheng2025fastf5tts}, which consistently outperforms uniform sampling in our experiments, as the baseline schedule. Teachers use 32
Sway-sampled steps with CFG.

\vspace{-5pt}

\subsection{Few-Step Synthesis}
\label{sec:main_results}

Table~\ref{tab:progression} shows that LFMD achieves strong few-step synthesis, approaching the 32-NFE teacher at three and four NFEs and even surpassing its WER, although SIM-o and UTMOS remain lower. Compared with IMF, LFMD is particularly effective at the most constrained budgets, with its largest gains at one and two NFEs, while both methods perform similarly at three and four NFEs. LFMD and IMF have essentially identical model-sampling RTF at matched size and NFE, while one-step inference is approximately $10\times$ faster than the corresponding 32-step teacher.

The main computational gain occurs during training. In our implementation, an LFMD update has a cost comparable to IMF with four teacher steps and is approximately 1.5, 2.5, and $4\times$ faster than IMF with 8, 16, and 32 teacher steps, respectively, although exact ratios depend on audio length. IMF is also highly sensitive to teacher trajectory discretization. Using only four teacher integration steps leads to poor low-NFE student performance, while increasing the integration budget does not fully resolve the issue under uniform discretization: even with 16 teacher steps, uniformly sampled trajectories remain substantially worse than Sway-sampled ones. At one student NFE, for example, a student trained from uniformly sampled teacher trajectories reaches approximately 35\% WER, compared with approximately 8\% when the teacher trajectories are Sway-sampled. We attribute this sensitivity partly to discretization error over long target intervals, which particularly affects one- and two-step students requiring the longest transports.

\vspace{-5pt}

\subsection{Scheduler Experiments}
\label{sec:schedules}

Across our experiments, TD-DP and O3S achieve broadly similar synthesis quality, with TD-DP performing slightly better at several operating points. The main advantage of TD-DP, however, is its simpler and more efficient schedule selection: unlike O3S, it requires neither auxiliary speech-model scoring nor repeated waveform generation, and its interval-cost graph is computed once and reused across NFE budgets rather than requiring a separate search for each budget. This reduces dependence on external evaluators and repeated budget-specific optimization.\vspace{-6pt}

\subsection{Fine-Tuning at Low NFE}
\label{sec:finetuning}

The four-step LFMD model achieves teacher-level WER at substantially lower target-generation cost and is naturally matched to the student. We therefore use a frozen copy as the self-distillation target, requiring only four evaluations instead of a 32-step teacher rollout.

At one NFE, frame-wise MSE refinement reduces Seed-TTS WER from 5.49\% to about 3.9\%, while soft-DTW further reduces it to 1.80\%, with corresponding improvements in SIM-o and UTMOS. Self-distillation also improves the two- and three-NFE models relative to their unrefined counterparts, reducing WER from 1.76\% to 1.68\% and from 1.72\% to 1.61\%, respectively. At one NFE, the refined model reaches 1.80\% WER, close to the 32-NFE teacher's 1.76\%, while maintaining strong SIM-o (.607) and UTMOS (3.41), although both remain further from the teacher (.667 and 3.76, respectively).

\vspace{-15pt}

\section{Conclusions}

We introduced LFMD, a memory-efficient adaptation of local flow-map distillation for few-step TTS that avoids numerical teacher trajectory integration during target construction. LFMD achieves strong low-NFE synthesis while reducing the training cost associated with trajectory-integrated targets. TD-DP further provides competitive schedules from a single reusable teacher-derived cost graph, without external audio-metric evaluation or budget-specific search. Finally, self-distillation from the strong four-step LFMD model improves lower-NFE generation, reaching 1.80\% WER at one NFE on Seed-TTS, close to the 32-NFE teacher's 1.76\%.

\section{Compliance with Ethical Standards}
This study uses publicly available speech datasets and did not involve the
collection of new human-subject data. No specific funding was received for
this study, and the authors declare no conflicts of interest.

\vspace{-15pt}

\bibliographystyle{IEEEbib}
\bibliography{references}

@misc{lin2026dsflow,
      title={{DSFlow}: Dual Supervision and Step-Aware Architecture for One-Step Flow Matching Speech Synthesis}, 
      author={Bin Lin and Peng Yang and Chao Yan and Xiaochen Liu and Wei Wang and Boyong Wu and Pengfei Tan and Xuerui Yang},
      year={2026},
      eprint={2602.09041},
      howpublished={arXiv preprint arXiv:2602.09041},
      archivePrefix={arXiv},
      primaryClass={cs.SD},
      url={https://arxiv.org/abs/2602.09041}, 
}

@misc{zheng2025fastf5tts,
      title={Accelerating Flow-Matching-Based Text-to-Speech via Empirically Pruned Step Sampling}, 
      author={Qixi Zheng and Yushen Chen and Zhikang Niu and Ziyang Ma and Xiaofei Wang and Kai Yu and Xie Chen},
      year={2025},
      eprint={2505.19931},
      howpublished={arXiv preprint arXiv:2505.19931},
      archivePrefix={arXiv},
      primaryClass={eess.AS},
      url={https://arxiv.org/abs/2505.19931}, 
}

@article{eskimez2024e2tts,
  title={{E2 TTS}: Embarrassingly Easy Fully Non-Autoregressive Zero-Shot {TTS}},
  author={Eskimez, Sefik Emre and Wang, Xiaofei and Thakker, Manthan and Li, Canrun and Tsai, Chung-Hsien and Xiao, Zhen and Yang, Hemin and Zhu, Zirun and Tang, Min and Tan, Xu and Liu, Yanqing and Zhao, Sheng and Kanda, Naoyuki},
  journal={arXiv preprint arXiv:2406.18009},
  year={2024}
}

@inproceedings{chen2025f5tts,
  title={F5-{TTS}: A Fairytaler that Fakes Fluent and Faithful Speech with Flow Matching},
  author={Chen, Yushen and Niu, Zhikang and Ma, Ziyang and Deng, Keqi and Wang, Chunhui and Zhao, Jian and Yu, Kai and Chen, Xie},
  booktitle={Proceedings of the 63rd Annual Meeting of the Association for Computational Linguistics (Volume 1: Long Papers)},
  pages={6255--6271},
  address={Vienna, Austria},
  publisher={Association for Computational Linguistics},
  doi={10.18653/v1/2025.acl-long.313},
  year={2025}
}

@article{boffi2024flowmapmatching,
  title={Flow Map Matching},
  author={Boffi, Nicholas M. and Albergo, Michael S. and Vanden-Eijnden, Eric},
  journal={arXiv preprint arXiv:2406.07507},
  year={2024}
}

@article{boffi2025flowmaps,
  title={How to Build a Consistency Model: Learning Flow Maps via Self-Distillation},
  author={Boffi, Nicholas M. and Albergo, Michael S. and Vanden-Eijnden, Eric},
  journal={arXiv preprint arXiv:2505.18825},
  year={2025}
}

@inproceedings{geng2025meanflow,
  title={Mean Flows for One-Step Generative Modeling},
  author={Geng, Zhengyang and Deng, Mingyang and Bai, Xingjian and Kolter, J. Zico and He, Kaiming},
  booktitle={Advances in Neural Information Processing Systems},
  year={2025}
}

@article{wang2025intmeanflow,
  title={{IntMeanFlow}: Few-Step Speech Generation with Integral Velocity Distillation},
  author={Wang, Wei and Cao, Rong and Guo, Yi and Chen, Zhengyang and Chen, Kuan and Huo, Yuanyuan},
  journal={arXiv preprint arXiv:2510.07979},
  year={2025}
}

@article{guo2025splitmeanflow,
  title={{SplitMeanFlow}: Interval Splitting Consistency in Few-Step Generative Modeling},
  author={Guo, Yi and Wang, Wei and Yuan, Zhihang and Cao, Rong and Chen, Kuan and Chen, Zhengyang and Huo, Yuanyuan and Zhang, Yang and Wang, Yuping and Liu, Shouda and Wang, Yuxuan},
  journal={arXiv preprint arXiv:2507.16884},
  year={2025}
}

@article{sabour2025alignyourflow,
  title={Align Your Flow: Scaling Continuous-Time Flow Map Distillation},
  author={Sabour, Amirmojtaba and Fidler, Sanja and Kreis, Karsten},
  journal={arXiv preprint arXiv:2506.14603},
  year={2025}
}

@inproceedings{elias2021paralleltacotron2,
  title={Parallel Tacotron 2: A Non-Autoregressive Neural TTS Model with Differentiable Duration Modeling},
  author={Elias, Isaac and Zen, Heiga and Shen, Jonathan and Zhang, Yu and Jia, Ye and Skerry-Ryan, R. J. and Wu, Yonghui},
  booktitle={Interspeech},
  pages={141--145},
  year={2021},
  doi={10.21437/Interspeech.2021-1461}
}

@inproceedings{yi2022softspeech,
  title={SoftSpeech: Unsupervised Duration Model in FastSpeech 2},
  author={Yi, Yuan-Hao and He, Lei and Pan, Shifeng and Wang, Xi and Zhang, Yuchao},
  booktitle={Interspeech},
  pages={1606--1610},
  year={2022},
  doi={10.21437/Interspeech.2022-887}
}

@inproceedings{song2023consistency,
  title={Consistency Models},
  author={Song, Yang and Dhariwal, Prafulla and Chen, Mark and Sutskever, Ilya},
  booktitle={Proceedings of the 40th International Conference on Machine Learning},
  year={2023}
}

@article{frans2024shortcut,
  title={One Step Diffusion via Shortcut Models},
  author={Frans, Kevin and Hafner, Danijar and Levine, Sergey and Abbeel, Pieter},
  journal={arXiv preprint arXiv:2410.12557},
  year={2024}
}

@article{anastassiou2024seedtts,
  title={{Seed-TTS}: A Family of High-Quality Versatile Speech Generation Models},
  author={Philip Anastassiou and Jiawei Chen and Jitong Chen and Yuanzhe Chen and Zhuo Chen and Ziyi Chen and Jian Cong and Lelai Deng and Chuang Ding and Lu Gao and Mingqing Gong and Peisong Huang and Qingqing Huang and Zhiying Huang and Yuanyuan Huo and Dongya Jia and Chumin Li and Feiya Li and Hui Li and Jiaxin Li and Xiaoyang Li and Xingxing Li and Lin Liu and Shouda Liu and Sichao Liu and Xudong Liu and Yuchen Liu and Zhengxi Liu and Lu Lu and Junjie Pan and Xin Wang and Yuping Wang and Yuxuan Wang and Zhen Wei and Jian Wu and Chao Yao and Yifeng Yang and Yuanhao Yi and Junteng Zhang and Qidi Zhang and Shuo Zhang and Wenjie Zhang and Yang Zhang and Zilin Zhao and Dejian Zhong and Xiaobin Zhuang},
  journal={arXiv preprint arXiv:2406.02430},
  year={2024},
  eprint={2406.02430},
  archivePrefix={arXiv},
  primaryClass={eess.AS},
  url={https://arxiv.org/abs/2406.02430}
}

@article{chen2022wavlm,
  title={{WavLM}: Large-Scale Self-Supervised Pre-Training for Full Stack Speech Processing},
  author={Chen, Sanyuan and Wang, Chengyi and Chen, Zhengyang and others},
  journal={arXiv preprint arXiv:2110.13900},
  year={2021},
  doi={10.1109/JSTSP.2022.3188113},
  url={https://arxiv.org/abs/2110.13900}
}

@article{chen2024gits,
  title={On the Trajectory Regularity of {ODE}-based Diffusion Sampling},
  author={Chen, Defang and Zhou, Zhenyu and Wang, Can and Shen, Chunhua and Lyu, Siwei},
  journal={arXiv preprint arXiv:2405.11326},
  year={2024},
  url={https://arxiv.org/abs/2405.11326}
}

@article{he2024emilia,
  title={{Emilia}: An Extensive, Multilingual, and Diverse Speech Dataset for Large-Scale Speech Generation},
  author={He, Haorui and Shang, Zengqiang and Wang, Chaoren and others},
  journal={arXiv preprint arXiv:2407.05361},
  year={2024},
  url={https://arxiv.org/abs/2407.05361}
}

@article{ho2022cfg,
  title={Classifier-Free Diffusion Guidance},
  author={Ho, Jonathan and Salimans, Tim},
  journal={arXiv preprint arXiv:2207.12598},
  year={2022},
  eprint={2207.12598},
  archivePrefix={arXiv},
  primaryClass={cs.LG},
  url={https://arxiv.org/abs/2207.12598}
}

@inproceedings{li2025dmospeech,
  title={{DMOSpeech}: Direct Metric Optimization via Distilled Diffusion Model in Zero-Shot Speech Synthesis},
  author={Yinghao Aaron Li and Rithesh Kumar and Zeyu Jin},
  booktitle={International Conference on Machine Learning},
  pages={35186--35208},
  series={Proceedings of Machine Learning Research},
  volume={267},
  year={2025},
  url={https://proceedings.mlr.press/v267/li25ay.html}
}

@article{li2026dmospeech2,
  title={{DMOSpeech 2}: Reinforcement Learning for Duration Prediction in Metric-Optimized Speech Synthesis},
  author={Li, Yinghao Aaron and Jiang, Xilin and Tao, Fei and Niu, Cheng and Xu, Kaifeng and Song, Juntong and Mesgarani, Nima},
  journal={arXiv preprint arXiv:2507.14988},
  year={2025},
  url={https://arxiv.org/abs/2507.14988}
}

@article{pei2025optimalsteps,
  title={Optimal Stepsize for Diffusion Sampling},
  author={Pei, Jianning and Hu, Han and Gu, Shuyang},
  journal={arXiv preprint arXiv:2503.21774},
  year={2025},
  url={https://arxiv.org/abs/2503.21774}
}

@article{radford2022whisper,
  title={Robust Speech Recognition via Large-Scale Weak Supervision},
  author={Radford, Alec and Kim, Jong Wook and Xu, Tao and Brockman, Greg and McLeavey, Christine and Sutskever, Ilya},
  journal={arXiv preprint arXiv:2212.04356},
  year={2022},
  url={https://arxiv.org/abs/2212.04356}
}

@article{saeki2022utmos,
  title={{UTMOS}: {UTokyo-SaruLab} System for {VoiceMOS} Challenge 2022},
  author={Saeki, Takaaki and Xin, Detai and Nakata, Wataru and Koriyama, Tomoki and Takamichi, Shinnosuke and Saruwatari, Hiroshi},
  journal={arXiv preprint arXiv:2204.02152},
  year={2022},
  url={https://arxiv.org/abs/2204.02152}
}

@article{siuzdak2023vocos,
  title={{Vocos}: Closing the gap between time-domain and {Fourier}-based neural vocoders for high-quality audio synthesis},
  author={Siuzdak, Hubert},
  journal={arXiv preprint arXiv:2306.00814},
  year={2023},
  url={https://arxiv.org/abs/2306.00814}
}

@inproceedings{yoon2025aptts,
  title={{APTTS}: Adversarial Post-training in Latent Flow Matching for Fast and High-fidelity Text-to-Speech},
  author={Yoon, Hyungchan and Lee, Chanwoo and Lee, Hoodong and Choi, Stanley Jungkyu},
  booktitle={Proc. Interspeech 2025},
  pages={5518--5522},
  year={2025},
  doi={10.21437/Interspeech.2025-455},
  url={https://www.isca-archive.org/interspeech_2025/yoon25_interspeech.html}
}

@article{zen2019libritts,
  title={{LibriTTS}: A Corpus Derived from {LibriSpeech} for Text-to-Speech},
  author={Zen, Heiga and Dang, Viet and Clark, Rob and Zhang, Yu and Weiss, Ron J. and Jia, Ye and Chen, Zhifeng and Wu, Yonghui},
  journal={arXiv preprint arXiv:1904.02882},
  year={2019},
  url={https://arxiv.org/abs/1904.02882}
}

@inproceedings{cuturi2017softdtw,
  title={{Soft-DTW}: a Differentiable Loss Function for Time-Series},
  author={Marco Cuturi and Mathieu Blondel},
  booktitle={International Conference on Machine Learning},
  pages={894--903},
  series={Proceedings of Machine Learning Research},
  volume={70},
  year={2017},
  url={https://proceedings.mlr.press/v70/cuturi17a.html}
}

@inproceedings{blondel2021dtw,
  title={Differentiable Divergences Between Time Series},
  author={Mathieu Blondel and Arthur Mensch and Jean-Philippe Vert},
  booktitle={International Conference on Artificial Intelligence and Statistics},
  pages={3853--3861},
  series={Proceedings of Machine Learning Research},
  volume={130},
  year={2021},
  url={https://proceedings.mlr.press/v130/blondel21a.html}
}

\end{document}